\documentclass[epj]{svjour}

\usepackage{epsfig}
\usepackage{graphicx}
\usepackage{color}

\begin{document}

\title{Luttinger theorem and imbalanced Fermi systems}
\author{P. Pieri\inst{1,2} \and G. Calvanese Strinati \inst{1,2}
}
\institute{School of Science and Technology, Physics Division, Universit\`{a} di Camerino,  62032 Camerino (MC), Italy
\and INFN, Sezione di Perugia, 06123 Perugia (PG), Italy}

\date{\today}

\abstract{
The proof of the Luttinger theorem, which was originally given for a normal Fermi liquid with equal spin populations formally described by the exact many-body theory at zero temperature, is here extended to an approximate theory given in terms of a ``conserving'' approximation also with spin imbalanced populations.
The need for this extended proof, whose underlying assumptions are here spelled out in detail, stems from the recent interest in superfluid trapped Fermi atoms with attractive inter-particle interaction,
for which the difference between two spin populations can be made large enough that superfluidity is destroyed and the system remains normal even at zero temperature.
In this context, we will demonstrate the validity of the Luttinger theorem separately for the two spin populations for any ``$\Phi$-derivable'' approximation, and illustrate it in particular for the self-consistent $t$-matrix approximation.
}
\maketitle

\section{Introduction}
The theory of normal Fermi liquids deals with a homogeneous system of interacting fermions in the normal phase close to the absolute zero of temperature in an essentially exact way, by expressing thermodynamic and dynamical quantities of interest in terms of a few phenomenological parameters \cite{AGD-1963,Nozieres-1964}.
These could, in principle, be calculated in terms of exact quantities of many-body theory, like the single-particle self-energy of the Dyson equation and the irreducible kernel of the 
two-particle Bethe-Salpeter equation \cite{AGD-1963,Nozieres-1964,Rickayzen-1980}.
In practice, however, to calculate these quantities approximations have unavoidably to be made.
The question then arises whether a given approximation may lead to unphysical violations of conservations laws or of important constraints.

In particular, one constraint that characterizes a Fermi liquid is the so-called Luttinger's sum rule (or Luttinger's theorem), which states that the volume enclosed by the Fermi surface 
of the interacting system is directly proportional to the particle density \cite{Luttinger-Ward-1960}.
And since the particle density is unaffected by the inter-particle interaction, the radius of the Fermi surface of the interacting system coincides with the Fermi wave vector of the non-interacting one \cite{Luttinger-1960}.
The proof of this theorem given in Refs. \cite{Luttinger-Ward-1960,Luttinger-1960}, as well as in the more recent works \cite{Oshikawa-2000,Praz-2005}, holds for the exact theory, and an important question is again what happens to it when approximations are adopted in realistic calculations. 

Approximations that are known to respect conservation laws are the so-called ``conserving'' approximations introduced by Baym and Kadanoff \cite{Baym-Kadanoff-1961,Baym-1962}, whereby certain classes of diagrams for the single-particle self-energy have to be taken together. 
These diagrams, in turn, contain the single-particle Green's functions which are self-consistently expressed in terms of the self-energy itself.
A sufficient condition to select these classes of diagrams is to generate the single-particle self-energy though a functional $\Phi$, by taking the functional derivative of $\Phi$ with respect to the
single-particle Green's function (``$\Phi$-derivable'' approximations) \cite{Baym-1962}.
In addition, one finds it stated that a $\Phi$-derivable approximation satisfies the Luttinger's sum rule \cite{Potthoff-2007}, since the original proof of this sum rule for the exact theory was also based on the existence of an exact functional $\Phi$ and can thus apply when an approximate form of $\Phi$ is introduced like in a $\Phi$-derivable approximation.

Recently, interest in Luttinger's theorem has arisen in the context of \emph{imbalanced} Fermi gases, in which the two spin components $\sigma = (\uparrow,\downarrow)$ have different densities $n_{\sigma}$.
Originally, this interest was stimulated by novel experimental studies of superfluid trapped Fermi atoms \cite{Ketterle-2006,Hulet-2006}, for which imbalanced populations can be maintained independently of the orbital degrees of freedom.
From a theoretical point of view, it turns out that the (non-self-consistent) $t$-matrix approximation (sometimes referred to as the $G_{0}G_{0}$ $t$-matrix \cite{PPSC-202}) or else its expanded NSR version \cite{NSR-1985}, which has often been used with (at least qualitative) success to describe the BCS-BEC crossover for gases with balanced populations, fails instead in the imbalanced case when two different chemical potentials $\mu_{\sigma}$ are introduced \cite{Hu-2006,Parish-2007,Ohashi-2013}.
For instance, if one calculates the densities for $\mu_{\uparrow} > \mu_{\downarrow}$, one finds inconsistently that $n_{\uparrow} < n_{\downarrow}$ in some regions of the phase diagram \cite{Hu-2006,Parish-2007}.
In addition, although the Luttinger's theorem does not hold for the superfluid phase which survives for moderate spin imbalance, its validity should be eventually restored for a strongly polarized Fermi gas, when the difference of the two spin populations is large enough that superfluidity is destroyed and the system remains normal even at zero temperature. 
Yet, the (non-self-consistent) $t$-matrix approximation with $n_{\uparrow} \ne n_{\downarrow}$ does not yield for the the radii of the two Fermi spheres the values one would expect by the Luttinger's theorem when applied separately to the two spheres \cite{Tartari-2011}.

A method to correct the failure of the NSR approach to fulfill the Luttinger's theorem for the imbalanced case was proposed in Ref. \cite{Urban-Schuck-2014}.
It consists in working with the zero-temperature form of the single-particle propagator $G_{0,\sigma}$ which contains from the outset the Fermi wave vector $k_{\rm F}^{\sigma}$ related to the density $n_{\sigma}$, instead of using the Matsubara form of $G_{0,\sigma}$ (in the zero-temperature limit) which instead contains the chemical potential $\mu_{\sigma}$ (and thus the associated wave vector 
$k_{\mu}^{\sigma}=\sqrt{2 m \mu_{\sigma}}$, where $m$ is the fermion mass and $\hbar =1$ troughout).

As a matter of fact, taking into account the difference between $k_{\mu}$ and $k_{\rm F}$ was found to be important also in the balanced case with $n_{\uparrow} = n_{\downarrow}$, for which 
the $G_{0}G_{0}$ $t$-matrix does not shows pathological behaviors.
It was, in fact, found in Ref. \cite{Camerino-JILA-2011} that the back-bending of the dispersions obtained from the single-particle spectral function $A(k,\omega)$ (with wave vector $k$ and frequency 
$\omega$) occurs at a wave vector $k_{\rm L}$ (referred there to as the Luttinger wave vector), which signals the presence of a remnant Fermi surface even in the superfluid phase.
It was further found in Ref. \cite{Camerino-JILA-2011} that $k_{\rm L}$ remains close to the Fermi wave vector $k_{\rm F}$ but departs markedly from $k_{\mu}$ over a wide coupling range, even approaching the molecular limit of the BCS-BEC crossover. 
[This was actually the reason for referring to $k_{\rm L}$ as the Luttinger wave vector, because the finding that the Fermi surface is (almost, in this case) unaffected by the interaction is reminiscent 
of the Luttinger's theorem for a Fermi liquid.]

The replacement of $k_{\mu}$ by $k_{\rm L}$ ($\simeq k_{\rm F}$) in the balanced superfluid case of Ref.\cite{Camerino-JILA-2011}, as well as the replacement of $k_{\mu}^{\sigma}$ by $k_{\rm F}^{\sigma}$ in the 
imbalanced normal case of Ref. \cite{Urban-Schuck-2014}, points to the need of introducing (especially for the imbalanced case) some sort of self-consistency in the $G_{0}G_{0}$ $t$-matrix, through an appropriate dressing of the bare single-particle propagator $G_{0}$ by interaction effects. 
The need to introduce at least a partial level of self-consistency in the $G_{0}G_{0}$ $t$-matrix was independently raised in Ref. \cite{Ohashi-2012}, where an extended $t$-matrix approach was introduced both for the balanced and the imbalanced case, which dresses the single-particle propagator $G_{0}$ closing the loop in the self-energy $\Sigma$.
For the imbalanced case, however, no work apparently exists that employs the fully self-consistent Green's function method (also called the $GG$ $t$-matrix or Luttinger-Ward 
method \cite{Haussmann-2007}), where all $G_{0}$ including that closing the loop in the self-energy $\Sigma$ are replaced by fully self-consistent $G$.
Among all the $t$-matrix approximations that have been considered, this is actually the only one to be conserving in the Baym-Kadanoff sense \cite{Baym-Kadanoff-1961,Baym-1962}.

For all the above reasons, although an extension of the proof of the Luttinger's theorem to conserving approximations and also for different spin populations may appear straightforward, 
we regard it both relevant and useful to provide here a schematic version of this proof.
This goes through the original Luttinger's line of arguments \cite{Luttinger-Ward-1960,Luttinger-1960} and emphasises the non-trivial assumptions underneath, having specifically in mind the self-consistent $GG$ $t$-matrix approximation for imbalanced Fermi systems.

\section{Proof of the Luttinger's theorem for conserving approximations in imbalanced Fermi systems}

We begin by considering the standard expression of the density for fermions with spin component $\sigma$ \cite{FW}
\begin{equation}
n_{\sigma} = \int \! \frac{d\mathbf{k}}{(2 \pi)^{3}} \frac{1}{\beta} \sum_{n} e^{i \omega_{n} \eta} \, \mathcal{G}_{\sigma}(\mathbf{k},\omega_{n})
\label{density-vs-single-particle-propagator}
\end{equation}
\noindent
where $\mathbf{k}$ is a wave vector, $\omega_{n}=(2n+1)\pi/\beta$ ($n$ integer) a fermionic Matsubara frequency, $\beta =(k_{B}T)^{-1}$ the inverse temperature ($k_{B}$ being the Boltzmann constant), and $\eta = 0^{+}$.
In this expression, the single-particle propagator
\begin{equation}
\mathcal{G}_{\sigma}(\mathbf{k},\omega_{n}) = \frac{1}{i\omega_{n} - \xi_{\mathbf{k}}^{\sigma} - \Sigma_{\sigma}(\mathbf{k},\omega_{n})} \, ,
\label{single-particle-propagator}
\end{equation}
\noindent
where $\xi_{\mathbf{k}}^{\sigma} = \varepsilon_{\mathbf{k}} - \mu_{\sigma}$ and $\varepsilon_{\mathbf{k}} = \mathbf{k}^{2}/(2m)$, contains in principle the full
self-energy $\Sigma_{\sigma}$ of the exact theory for given spin component.
Although we have chosen to work with the Matsubara formalism so as to introduce the chemical potentials $\mu_{\sigma}$ at the outset, in the following we shall take the zero-temperature limit in such 
a way that the Matsubara frequencies $\omega_{n}$ are densely distributed and one can replace accordingly:
\begin{equation}
\frac{1}{\beta} \sum_{n} \,\, \longrightarrow \,\,  \int_{-\infty}^{+\infty} \! \frac{d\omega}{2 \pi} \,\, .
\label{sum-into-integral}
\end{equation}

Following Ref.~\cite{AGD-1963}, we next perform the following manipulations on Eq.(\ref{density-vs-single-particle-propagator}).
We take the logarithm of $\mathcal{G}_{\sigma}(\mathbf{k},\omega_{n})$ and then the derivative of the resulting expression with respect to $i\omega_{n}$, to obtain
\begin{eqnarray}
\frac{\partial}{\partial \, i \omega_{n}} \ln \mathcal{G}_{\sigma}(\mathbf{k},\omega_{n}) = - 
\frac{ 1 \, - \, \frac{\partial}{\partial \, i \omega_{n}} \Sigma_{\sigma}(\mathbf{k},\omega_{n})}{i\omega_{n} - \xi_{\mathbf{k}}^{\sigma} - \Sigma_{\sigma}(\mathbf{k},\omega_{n})} &&
\nonumber \\
=
- \mathcal{G}_{\sigma}(\mathbf{k},\omega_{n}) \, \left[ 1 \, - \, \frac{\partial}{\partial \, i \omega_{n}} \Sigma_{\sigma}(\mathbf{k},\omega_{n}) \right]&& ,
\label{partial-derivative}
\end{eqnarray}
in such a way that Eq.(\ref{density-vs-single-particle-propagator}) can be rewritten in the form:
\begin{eqnarray}
n_{\sigma} = \int \! \frac{d\mathbf{k}}{(2 \pi)^{3}} \frac{1}{\beta} \sum_{n} e^{i \omega_{n} \eta}
\left[ - \frac{\partial}{\partial \, i \omega_{n}} \ln \mathcal{G}_{\sigma}(\mathbf{k},\omega_{n})\right.&&\nonumber\\ 
+ \left.\mathcal{G}_{\sigma}(\mathbf{k},\omega_{n}) \frac{\partial}{\partial \, i \omega_{n}} \Sigma_{\sigma}(\mathbf{k},\omega_{n}) \right]&& .
\label{density-vs-single-particle-propagator-modified}
\end{eqnarray}
\noindent
The point is now to show that the second term within brackets on the right-hand side of Eq.(\ref{density-vs-single-particle-propagator-modified}) gives a vanishing  contribution to $n_{\sigma} $,
and this not only for the exact (Fermi liquid) theory but also for any approximate theory for $\Sigma_{\sigma}$ (and thus for $\mathcal{G}_{\sigma}$) which is $\Phi$-derivable in the Baym-Kadanoff sense \cite{Baym-Kadanoff-1961,Baym-1962}.
For an imbalanced system, one is then assuming that there exists a functional $\Phi_{\sigma}$ of $\mathcal{G}_{\uparrow}$ and $\mathcal{G}_{\downarrow}$ for both spin components, such that $\Sigma_{\sigma}$ can be obtained from a functional derivative in the form \cite{Baym-1962}:
\begin{equation}
\Sigma_{\sigma}(x_{1},x_{2}) = \frac{\delta \Phi_{\sigma}}{\delta \mathcal{G}_{\sigma}(x_{2},x_{1})}
\label{functional-derivative}
\end{equation}
\noindent
where $x=(\mathbf{r},\tau)$ contains the space variable $\mathbf{r}$ and the imaginary time $\tau$.
A variation $\mathcal{G}_{\sigma} \rightarrow \mathcal{G}_{\sigma} + \delta \mathcal{G}_{\sigma}$ thus entails the following variation in $ \Phi$:
\begin{eqnarray}
\delta \Phi_{\sigma} & = & \int \! dx_{1} dx_{2} \, \frac{\delta \Phi_{\sigma}}{\delta \mathcal{G}_{\sigma}(x_{2},x_{1})} \, \delta \mathcal{G}_{\sigma}(x_{2},x_{1})
\nonumber \\
                 &=& \int \! dx_{1} dx_{2} \, \Sigma_{\sigma}(x_{1},x_{2}) \, \delta \mathcal{G}_{\sigma}(x_{2},x_{1}) 
\nonumber \\
& = & V \beta \int \! \frac{d\mathbf{k}}{(2 \pi)^{3}} \frac{1}{\beta} \sum_{n} \, \Sigma_{\sigma}(\mathbf{k},\omega_{n}) \, \delta \mathcal{G}_{\sigma}(\mathbf{k},\omega_{n})                 
\label{variation-Phi}
\end{eqnarray}
\noindent
where $V$ is the volume of the system.
Specifically, in each diagram making up $\Phi_{\sigma}$ one performs a variation of all $\mathcal{G}_{\sigma}$ that enter the diagram, by shifting their frequency argument 
$i \omega \rightarrow i \omega + i \delta \omega_{0}$ in the zero-temperature limit, while keeping unchanged all $\mathcal{G}_{\bar{\sigma}}$.
One then claims that $\Phi_{\sigma}$ is left unchanged by this variation, in such a way that:
\begin{eqnarray}
\frac{\partial \tilde{\Phi}_{\sigma}}{\partial i \omega} &=& \int \! \frac{d\mathbf{k}}{(2 \pi)^{3}} \, \int_{-\infty}^{+\infty} \! \frac{d\omega}{2 \pi} \,\,
\Sigma_{\sigma}(\mathbf{k},\omega) \, \frac{\partial \mathcal{G}_{\sigma}(\mathbf{k},\omega)}{\partial i \omega} 
\nonumber \\
 &=& \, 0 
\label{particular-variation-Phi}
\end{eqnarray}
\noindent
where $\tilde{\Phi}_{\sigma} = \Phi_{\sigma} / (V \beta)$.

\begin{figure}
\begin{center}
\includegraphics[width=8.5cm,angle=0]{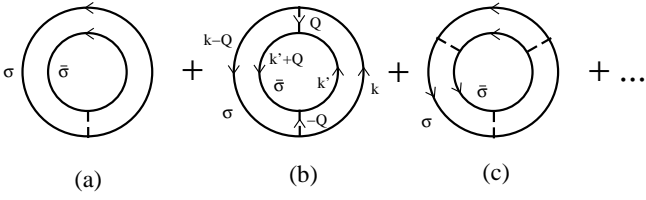}
\caption{First few diagrammatic terms corresponding to the functional $\Phi$ associated with the $GG$ $t$-matrix approximation for an imbalanced Fermi system. 
              Full lines denote the self-consistent propagators $\mathcal{G}_{\sigma}$ and broken lines the inter-particle interaction $v$ (taken of the contact type, 
              for which fermions of spin ${\sigma}$ interact only with fermions of opposite spin $\bar{\sigma}$).}
\label{Figure-1}
\end{center}
\end{figure}

In particular, this property can be shown to hold for the $GG$ $t$-matrix approximation for an imbalanced Fermi system, for which a few diagrams that correspond to the functional $\tilde{\Phi}_{\sigma}$ are shown in Fig.~\ref{Figure-1}. 
[Note that $\tilde{\Phi}_{\sigma}$ is symmetric under the interchange $\sigma \leftrightarrow \bar{\sigma}$ and has thus the same value for both spin species.]
As an example, let's consider the second-order diagram (b) of Fig.~\ref{Figure-1} which contains two interaction lines.
With the short-hand four-vector notation $k=(\mathbf{k},\omega_{n})$ and $Q=(\mathbf{Q},\Omega_{\nu})$, where $\Omega_{\nu}=2\pi\nu/\beta$ ($\nu$ integer) is a bosonic Matsubara frequency (also considered in the zero-temperature limit), as well as with the summation notation
\begin{equation}
\sum \limits_{k}^{} \, \longleftrightarrow \, \int \! \frac{d\mathbf{k}}{(2 \pi)^{3}} \, \int_{-\infty}^{+\infty} \! \frac{d\omega}{2 \pi} \, ,
\label{short-hand-notation}
\end{equation}
\noindent
we write for the contribution $\tilde{\Phi}_{\sigma}^{(b)}$ to the functional $\tilde{\Phi}_{\sigma}$ from this diagram:
\begin{eqnarray}
&-&\tilde{\Phi}_{\sigma}^{(b)} = \frac{1}{2} \!\sum \limits_{k,k',Q}^{}\!\!\!v(Q) v(-Q) \mathcal{G}_{\sigma}(k)  \mathcal{G}_{\sigma}(k-Q) \mathcal{G}_{\bar{\sigma}}(k')  \mathcal{G}_{\bar{\sigma}}(k'+Q)
\nonumber \\
& = & \frac{1}{2} \sum \limits_{\tilde{k},k',Q}^{} v(Q) v(-Q) \mathcal{G}_{\sigma}(\tilde{k}+\delta\omega_0)  \mathcal{G}_{\sigma}(\tilde{k}+\delta\omega_0 -Q)\mathcal{G}_{\bar{\sigma}}(k')  \nonumber\\ 
&&\,\,\,\,\,\, \,\,\,\,\,\,\,\,\times\mathcal{G}_{\bar{\sigma}}(k'+Q)
\nonumber \\
& = & - \tilde{\Phi}_{\sigma}^{(b)} + \frac{1}{2} \sum \limits_{\tilde{k},k',Q}^{} v(Q) v(-Q) 
\left[ \frac{\partial \mathcal{G}_{\sigma}(\tilde{k})}{\partial i \tilde{\omega}}  \mathcal{G}_{\sigma}(\tilde{k}-Q)\mathcal{G}_{\bar{\sigma}}(k')\right. \nonumber\\
&\times&\left.\mathcal{G}_{\bar{\sigma}}(k'+Q) 
+ \mathcal{G}_{\sigma}(\tilde{k}) \frac{\partial \mathcal{G}_{\sigma}(\tilde{k}-Q)}{\partial i \tilde{\omega}} \mathcal{G}_{\bar{\sigma}}(k')  \mathcal{G}_{\bar{\sigma}}(k'+Q)\right] i \delta \omega_{0} 
\nonumber \\
& = & - \tilde{\Phi}_{\sigma}^{(b)} - \left[ \sum \limits_{\tilde{k}}^{} \frac{\partial \mathcal{G}_{\sigma}(\tilde{k})}{\partial i \tilde{\omega}} \, \Sigma_{\sigma}(\tilde{k}) \right] i \delta \omega_{0}
\label{functional-Phi-example}
\end{eqnarray}

\noindent
to first order in $\delta \omega_{0}$, where the self-energy $\Sigma_{\sigma}$ on the right-hand side of Eq.(\ref{functional-Phi-example}) corresponds to the contribution of this particular diagram.
[Without loss of generality, we have here assumed for simplicity that only opposite-spin fermions interact with each other, as it is the case for a contact potential.]
The identity (\ref{particular-variation-Phi}) thus holds for this particular diagram, and independently for each spin component.
This proof can be readily extended to all other diagrams of Fig.~\ref{Figure-1}, which all together are associated with the $GG$ $t$-matrix approximation of interest \cite{footnote-1}.

\begin{figure*}
\begin{center}
\includegraphics[width=10.0cm,angle=0]{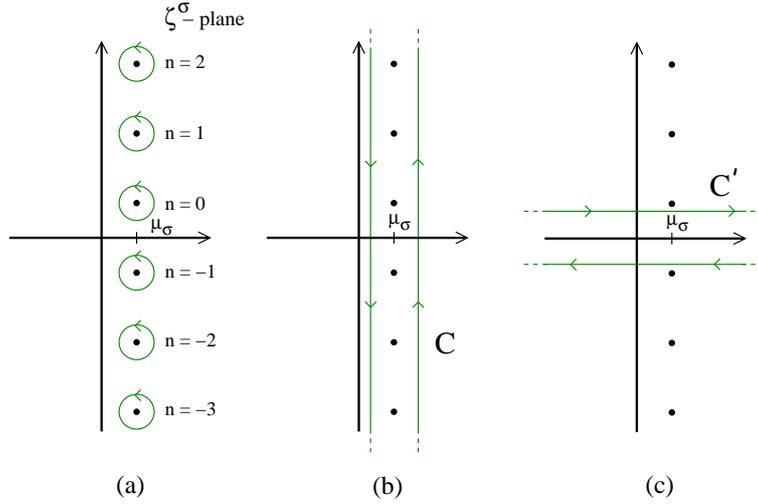}
\caption{Contours for evaluating the frequency sum of Eq.(\ref{density-reduced-expression-2}) in the complex $\zeta^{\sigma}$-plane.
The integration around the small circles centred at $\mu_{\sigma} + i \omega_{n}$ with $\omega_{n}=(2n+1)\pi/\beta$ ($n=0,\pm 1, \pm 2, \cdots$) in (a), is first transformed into
an integration along the contour $C$ that runs parallel to the imaginary axis in (b), and then into an integration along the contour $C'$ that runs parallel to the real axis in (c).}
\label{Figure-2}
\end{center}
\end{figure*} 
More generally, the result (\ref{particular-variation-Phi}), on which the present derivation of the Luttinger theorem for imbalanced systems is based, remains valid for \emph{any} $\Phi$-derivable approximation provided that the interaction between fermions does not produce spin flips. 
This is because any approximate form of $\Phi_{\sigma}$ contains sets of diagrams where closed loops of Green's functions of a given species are mutually connected by interaction lines.
In each of these loops that correspond to the same spin species, one can single out a common fermionic frequency integrated from $-\infty$ to $+\infty$, such that a constant shift $\delta \omega_0$ does not alter the value of $\Phi_{\sigma}$, thus implying that the equation (\ref{particular-variation-Phi}) is verified \cite{footnote-2}.  

With the help of the result (\ref{particular-variation-Phi}), the second term on the right-hand side of the expression (\ref{density-vs-single-particle-propagator-modified}) for the density can be manipulated as follows via an integration by parts in the zero-temperature limit:
\begin{eqnarray}
 \int \! \frac{d\mathbf{k}}{(2 \pi)^{3}} \, \int_{-\infty}^{+\infty} \! \frac{d\omega}{2 \pi} \, 
e^{i \omega \eta} \, \mathcal{G}_{\sigma}(\mathbf{k},\omega) \, \frac{\partial \Sigma_{\sigma}(\mathbf{k},\omega)}{\partial i \omega}&& 
\nonumber \\
=  \frac{1}{2 \pi i} \int \! \frac{d\mathbf{k}}{(2 \pi)^{3}} \, \mathcal{G}_{\sigma}(\mathbf{k},\omega) \, \Sigma_{\sigma}(\mathbf{k},\omega) \vert_{\omega=-\infty}^{\omega=+\infty}&&
\nonumber \\
 -  \int \! \frac{d\mathbf{k}}{(2 \pi)^{3}} \, \int_{-\infty}^{+\infty} \! \frac{d\omega}{2 \pi} \, \Sigma_{\sigma}(\mathbf{k},\omega) \frac{\partial \mathcal{G}_{\sigma}}{\partial \, i \omega} (\mathbf{k},\omega)&&
 \nonumber \\
= 0 \phantom{aaaaaaaaaaaaaaaaaaaaaaaaaaaaaa}&&
\label{manipulation}
\end{eqnarray}
\noindent
since not only the second term but also the first term of the right-hand side of Eq.(\ref{manipulation}) vanishes owing to the property $ \mathcal{G}_{\sigma}(k) \Sigma_{\sigma}(k) \rightarrow 0$ when $|\omega| \rightarrow \infty$.
In conclusion, the expression (\ref{density-vs-single-particle-propagator-modified}) reduces to the form:
\begin{equation}
n_{\sigma} = - \int \! \frac{d\mathbf{k}}{(2 \pi)^{3}} \, \frac{1}{\beta} \sum_{n}  \, e^{i \omega_{n} \eta} \, \frac{\partial}{\partial \, i \omega_{n}} \ln \mathcal{G}_{\sigma}(\mathbf{k},\omega_{n})
\label{density-reduced-expression-1}
\end{equation}
\noindent
where, for convenience, we have restored the finite temperature notation. 
It is convenient at this point to follow the arguments of Refs.~\cite{Luttinger-Ward-1960,Luttinger-1960} and introduce the variable $\zeta_{n}^{\sigma} = i \omega_{n} + \mu_{\sigma}$, such that
$\mathcal{G}_{\sigma}(\mathbf{k},\omega_{n}) = [\zeta_{n}^{\sigma} - \varepsilon_{\mathbf{k}} - \Sigma_{\sigma}(\mathbf{k},\omega_{n})]^{-1}$.
Equation (\ref{density-reduced-expression-1}) then becomes:
\begin{equation}
n_{\sigma} = \int \! \frac{d\mathbf{k}}{(2 \pi)^{3}} \, \frac{1}{\beta} \sum_{n} \, e^{\zeta_{n}^{\sigma} \eta} \, \frac{\partial}{\partial \, \zeta_{n}^{\sigma}} 
\ln [\varepsilon_{\mathbf{k}} + \Sigma_{\sigma}(\mathbf{k},\zeta_{n}^{\sigma}) - \zeta_{n}^{\sigma}]
\label{density-reduced-expression-2}
\end{equation}
\noindent
since $\ln(-z) = \ln(z) \pm i \pi$.
The sum over $n$ in Eq.(\ref{density-reduced-expression-2}) can be transformed in the usual way into an integral over the complex variable $\zeta^{\sigma}$, by recourse to the function 
$f(\zeta^{\sigma})\equiv(-\beta)[e^{\beta(\zeta^{\sigma}-\mu_\sigma)}+1]^{-1}$ which has simple poles with unit residue at $ \zeta_{n}^{\sigma} = i \omega_{n} + \mu_{\sigma}$ [cf. Fig.~\ref{Figure-2}(a)].
One thus introduces the contour $C$ of Fig.~\ref{Figure-2}(b) which runs parallel to the imaginary $\zeta^{\sigma}$-axis,
and then deforms it into the contour $C'$ of Fig.~\ref{Figure-2}(c) which runs just above and below the real $\zeta^{\sigma}$-axis, to take into account the singularities of the single-particle propagator
$\mathcal{G}_{\sigma}(\mathbf{k},\zeta^{\sigma})$ across the real $\zeta^{\sigma}$-axis (once this is obtained from $\mathcal{G}_{\sigma}(\mathbf{k},\zeta_{n}^{\sigma})$ through analytic continuation).
This is possible because the retarded (advanced) single-particle propagator $\mathcal{G}_{\sigma}^{(R)}$ ($\mathcal{G}_{\sigma}^{(A)}$) has no singularities in the upper (lower) half $\zeta^{\sigma}$-plane, such that $\varepsilon_{\mathbf{k}} + \Sigma_{\sigma}^{(R)}(\mathbf{k},\zeta^{\sigma}) - \zeta^{\sigma}$ has no zero in the upper half plane and 
$\varepsilon_{\mathbf{k}} + \Sigma_{\sigma}^{(A)}(\mathbf{k},\zeta^{\sigma}) - \zeta^{\sigma}$ has no zero in the lower half plane.
We can then write for the frequency sum in Eq.(\ref{density-reduced-expression-2}):
\begin{eqnarray}
& & \frac{1}{\beta} \sum_{n} \, e^{\zeta_{n}^{\sigma} \eta} \, \frac{\partial}{\partial \, \zeta_{n}^{\sigma}} \ln [\varepsilon_{\mathbf{k}} + \Sigma_{\sigma}(\mathbf{k},\zeta_{n}^{\sigma}) - \zeta_{n}^{\sigma}]
\nonumber \\
& = & \frac{1}{\beta} \int_{C} \! \frac{d \zeta^{\sigma}}{2 \pi i} e^{\zeta^{\sigma}\eta} \left( \frac{\partial}{\partial \, \zeta^{\sigma}} \ln [\varepsilon_{\mathbf{k}} + \Sigma_{\sigma}(\mathbf{k},\zeta^{\sigma}) - \zeta^{\sigma}] \right)f(\zeta^{\sigma})
\nonumber \\
& = & -\frac{1}{\beta} \int_{C'} \! \frac{d \zeta^{\sigma}}{2 \pi i} \, e^{\zeta^{\sigma}\eta} \ln [\varepsilon_{\mathbf{k}} + \Sigma_{\sigma}(\mathbf{k},\zeta^{\sigma}) - \zeta^{\sigma}] 
\frac{\partial f(\zeta^{\sigma})}{\partial \, \zeta^{\sigma}} 
\nonumber \\
& = & \frac{1}{2 \pi i} \left\{ - \ln [\varepsilon_{\mathbf{k}} + \Re \Sigma_{\sigma}(\mathbf{k},\zeta^{\sigma}=\mu_{\sigma}) - (\mu_{\sigma} + i \eta)] \right.
\nonumber \\
& & \left. \hspace{1.0cm} + \ln [\varepsilon_{\mathbf{k}} + \Re \Sigma_{\sigma}(\mathbf{k},\zeta^{\sigma}=\mu_{\sigma}) - (\mu_{\sigma} - i \eta)] \right\} 
\nonumber \\
& = & \Theta \left (\mu_{\sigma} - \varepsilon_{\mathbf{k}} - \Re \Sigma_{\sigma}(\mathbf{k},\zeta^{\sigma}=\mu_{\sigma}) \right) \, .
\label{partial-density-reduced-expression}
\end{eqnarray}
\noindent
Note that to obtain the last line of Eq.(\ref{partial-density-reduced-expression}) that holds for \emph{any} value of $\mathbf{k}$, we have:
(i) used the relation $- \frac{\partial}{\partial \, z} \frac{1}{e^{\beta (z - \mu)}+1} \\ = \delta(z - \mu)$ for real $z$ in the $T \rightarrow 0$ limit;
(ii) made use of the property $\Im \Sigma_{\sigma}(\mathbf{k},\zeta^{\sigma}) < 0$ ($> 0$) just above (below) the real axis;
(iii) replaced $\Im \Sigma_{\sigma}(\mathbf{k},\zeta^{\sigma}) \rightarrow \mp \eta = \mp 0^{+}$ in the argument of the logarithm, to the extent that
$\Im \Sigma_{\sigma}(\mathbf{k},\zeta^{\sigma}=\mu_{\sigma})=0$ for a Fermi liquid \cite{footnote-3};
(iv) used the property $\lim_{\eta \rightarrow 0} \ln (a \pm i \eta) = \ln|a| \pm i \pi \Theta(-a)$ that holds on the principal branch of the logarithm for any real number $a$, $\Theta$ being the unit step function.

Entering the result (\ref{partial-density-reduced-expression}) into Eq.(\ref{density-reduced-expression-2}), we obtain eventually
\begin{equation}
n_{\sigma} = \int \! \frac{d\mathbf{k}}{(2 \pi)^{3}} \, \Theta \left (\mu_{\sigma} - \varepsilon_{\mathbf{k}} - \Re \Sigma_{\sigma}(\mathbf{k},\zeta^{\sigma}=\mu_{\sigma}) \right) \, ,
\label{density-expression-final}
\end{equation}
\noindent
which shows that the total volume of $\mathbf{k}$-space contributing to the particle density remains the same of the non-interacting system, to the extent that the mean particle density is unaffected by the 
inter-particle interaction.
For an isotropic system, the further assumption that the quantity $\varepsilon_{\mathbf{k}} + \Re \Sigma_{\sigma}(\mathbf{k},\zeta^{\sigma}=\mu_{\sigma}) - \mu_{\sigma}$ is an increasing function of
$|\mathbf{k}|$ and that  $\varepsilon_{\mathbf{k}} + \Re \Sigma_{\sigma}(\mathbf{k},\zeta^{\sigma}=\mu_{\sigma}) - \mu_{\sigma} = 0$ has a single solution
for given $\sigma$, identifies a special value of $|\mathbf{k}|$ (say, $k_{\rm L}^{\sigma}$) associated with the \emph{interacting Fermi surface}. 
This surface bounds the region of $\mathbf{k}$-space giving a non-vanishing contribution to the integral in Eq.(\ref{density-expression-final}), in such a way that:
\begin{equation}
n_{\sigma} = \frac{1}{(2 \pi)^{3}} \, \frac{4 \pi}{3} \, \left( k_{\rm L}^{\sigma} \right)^{3} \, .
\label{density-vs-Luttinger_wavevector}
\end{equation}
\noindent
The result (\ref{density-vs-Luttinger_wavevector}) should be compared with the expression 
$n_{\sigma} = \frac{1}{(2 \pi)^{3}} \frac{4 \pi}{3}  \left( k_{\rm F}^{\sigma} \right)^{3}$ for the non-interacting Fermi system with $\Sigma_{\sigma}=0$.
This yields the desired result
\begin{equation}
k_{\rm L}^{\sigma} = k_{\rm F}^{\sigma}
\label{Luttinger_theorem}
\end{equation}
\noindent
known as the \emph{Luttinger theorem}, which states that the radius of the Fermi surface of the interacting system coincides with that of the non-interacting system for each $\sigma$-species.
This completes our proof.

\section{Concluding remarks}

In this paper, we have extended the proof of the Luttinger theorem, that was originally conceived for the \emph{exact} theory of a normal Fermi liquid, to any \emph{approximate} theory based on a $\Phi$-derivable (conserving) approximation also for the case of different spin populations.
In this context, we have been concerned, in particular, with the self-consistent $t$-matrix approximation that can be used to describe a superfluid Fermi system with an attractive inter-particle interaction
throughout the BCS-BEC crossover.
In this case, the Luttinger theorem becomes relevant when the imbalance between the spin populations is large enough that the system becomes normal even at zero temperature.
In the process, we have pointed out a number of assumptions that have to be verified by the approximate theory for the Luttinger theorem to hold separately for the spin populations, and we have also identified at which stage of the proof the non-self-consistent version of the $t$-matrix approximation fails to satisfy the required assumptions.

Finally, it is worth pointing out that, although the self-consistent $t$-matrix approximation, to which the Luttinger theorem (\ref{Luttinger_theorem}) applies, amounts to a truncation of the expansion of the functional $\Phi$ since it sums up only a specific subclass of skeleton diagrams, by no means can this approximation be considered a ``weak-coupling" approximation in the standard sense \cite{Potthoff-2007}.
This is because, already at the level of its non-self-consistent version, in the balanced case the $t$-matrix approximation can account for the physics of the BCS-BEC crossover at finite temperature \cite{Strinati-2012}, whereby the system evolves from the BCS limit of a weak inter-particle attraction when Cooper pairs are highly overlapping, to the BEC limit of a strong 
inter-particle attraction when composite bosons are not overlapping.
And also in the imbalanced case, the non-self-consistent $t$-matrix approximation yields the correct result (as compared with Monte-Carlo calculations) in the strong-coupling limit of the inter-particle interaction, when considering the extreme imbalanced situation of a single spin-$\downarrow$ fermion embedded in a sea of spin-$\uparrow$ fermions \cite{Combescot-2008}.\\

\noindent
Partial support from Italian MIUR through the PRIN 2015 program (Contract No. 2015C5SEJJ001) is acknowledged.

\vspace{0.5cm}
{\bf Author contribution statement} \\

\noindent
G.C.S. conceived the work. Both authors contributed to the proof of the theorem and to the writing of the manuscript.  



\begin{thebibliography}{99}
\bibitem{footnote-2} On physical grounds, the functional $\Phi_{\sigma}$ should remain symmetric under the interchange $\sigma \leftrightarrow \bar{\sigma}$ whatever the $\Phi$-derivable approximation,
                                 although this property is not relevant for the present proof of the Luttinger theorem.
\end{thebibliography}
\end{document}